\documentclass[mathleft,fleqn,%
]{an}
\usepackage{graphicx}
\usepackage[varg]{txfonts}
\begin{document}

\Pagespan{1}{}
\Yearpublication{2014}%
\Yearsubmission{2014}%
\Month{0}%
\Volume{999}%
\Issue{0}%
\DOI{asna.201400000}%

\title{Affordable echelle spectroscopy of the eccentric HAT-P-2, WASP-14 and XO-3 planetary systems with a sub-meter-class telescope}

\author{Z. Garai\inst{1}\fnmsep\thanks{Corresponding author:
        {zgarai@ta3.sk}}
\and  T. Pribulla\inst{1}
\and  \v{L}. Hamb\'alek\inst{1}
\and  E. Kundra\inst{1}
\and  M. Va\v{n}ko\inst{1}
\and  S. Raetz\inst{2}\fnmsep\inst{3}
\and  M. Seeliger\inst{2}
\and  C. Marka\inst{4}
\and  H. Gilbert\inst{2}
}
\titlerunning{Affordable echelle spectroscopy of eccentric planetary systems}
\authorrunning{Z. Garai et al.}
\institute{
Astronomical Institute, Slovak Academy of Sciences, 059 60 Tatransk\'a Lomnica, Slovakia
\and
Astrophysical Institute and University Observatory, F. Schiller University, Schillergaesschen 2-3, 07745 Jena,
Germany
\and
European Space Agency, Keplerlaan 1, 2200 AG, Noordwijk, The Netherlands
\and
Institute for Milimetric Radioastronomy, Avenida Divina Pastora 7, 18012 Granada, Spain 
}

\received{XXXX}
\accepted{XXXX}
\publonline{XXXX}


\abstract{%
A new off-shelf low-cost echelle spectrograph was installed recently on the 0.6m telescope at the Star\'a Lesn\'a Observatory (Slovakia). In this paper we describe in details the radial velocity (RV) analysis of the first three transiting planetary systems, HAT-P-2, WASP-14 and XO-3, observed with this instrument. Furthermore, we compare our data with the RV data achieved with echelle spectrographs of other sub-meter-, meter- and two-meter-class telescopes in terms of their precision. Finally, we investigate the applicability of our RV data for modeling orbital parameters.    
}

\maketitle


\section{Introduction}
The radial velocity (hereafter RV) method (see e.g., Mayor \& Queloz 1995; Vogt et al. 2010, 2014, 2015 and references therein) was the 1st successful extrasolar planet detection method. As of January 1, 2016, there were more than 630 confirmed exoplanets in 470 planetary systems (110 multiple) detected by RV measurements\footnote{see {\tt www.exoplanet.eu}}. This method, along with the exoplanet transit method (see e.g., Charbonneau et al. 2000; Borucki et al. 2011, 2013; Batalha et al. 2013 or Southworth 2012 and references therein), provides a wealth of information about a planetary system. 

A disadvantage of the RV method has been the very expensive instrumentation elements, mainly due to the high costs of stable high-resolution spectrographs and the requirements on the telescope diameter. Recent advances in instrumentation (e.g., the fiber-fed technique enables to locate the spectrograph in a thermally isolated room or box), however, enable affordable Doppler planet detections even with sub-meter-class telescopes. During the last decade, several sub-meter- and meter-class telescopes have been equipped with an echelle spectrograph: e.g. (1) in India, Mt Abu -- a 1.2m telescope (Chakraborty, Mahadevan \& Richardson 2008), (2) in Hungary, Piszk\'estet\H{o} and Szombathely -- a 1.0m and 0.5m telescope, respectively (Cs\'ak et al. 2014), (3) in Germany, Gro{\ss}schwabhausen -- a 0.6/0.9m telescope (Mugrauer, Avila \& Guirao 2014) and (4) in Slovakia, Star\'a Lesn\'a -- a 0.6m telescope (Pribulla et al. 2015).  

The Star\'a Lesn\'a Observatory in Slovakia (20$\degr$17'21" E, 49$\degr$09'06" N, 810 meters a.s.l. for the G1 observing pavilion) is equipped with two 0.6m (f/12.5) Zeiss reflecting telescopes located in the G1 and G2 observing pavilions\footnote{see {\tt http://www.astro.sk/l3.php?p3=sto}}. In March 2013 a new fiber-fed echelle spectrograph was procured from the French company Shelyak\footnote{see {\tt www.shelyak.com}} and mounted at the G1 pavilion in the Cassegrain focus of the telescope. Several known transiting and non-transiting exoplanet systems were already observed during the years 2013 and 2014 in G1 with this spectrograph, in order to assess the RV stability and precision which can be obtained. First RV measurements of $\tau$ Boo, HAT-P-2 and WASP-14 systems were described by Pribulla et al. (2015). In this paper we describe in details the RV analysis of the first three transiting planetary systems, HAT-P-2, WASP-14 and XO-3, observed at the Star\'a Lesn\'a Observatory. The objects were selected according to the RV amplitude, brightness and the sky position. All three systems are, however, very interesting, because they are characterized by close-in, but apparently eccentric orbits, and therefore represent potentially important systems to constrain the migration, tidal and thermal evolution of gas giant planets (Bakos et al. 2007; Joshi et al. 2009; Johns-Krull et al. 2008). 

In comparison with the paper published by Pribulla et al. (2015), we used more RV observations achieved at the Star\'a Lesn\'a Observatory and we derived spectroscopic orbits using these RV data. The scientific goal of this paper is, however, not to improve parameters of these planetary systems, rather to compare RV observations from the Star\'a Lesn\'a Observatory with other RV data and to validate the applicability of our RV data for modeling orbital parameters. First, we reduced and analyzed our RV observations. Subsequently, we compared our data with previously published RV data (Cs\'ak et al. 2014; Joshi et al. 2009; H\'ebrard et al. 2008). We were curious about the precision of our measurements in comparison to the RV data achieved with echelle spectrographs of other sub-meter-, meter- and two-meter-class telescopes. Another question was the applicability of our RV data for modeling orbital parameters. For this purpose the previously published data were analyzed in the same way as our RV data in order to determine and compare parameters. Finally, we combined and analyzed all used RV data per object.

The layout of the paper is as follows. In Sect. 2, we briefly describe our instrumentation. In Sect. 3, we present the observations and data analysis. Sect. 4 describes our results. Our conclusions are discussed in Sect. 5.                      


\section{Properties of the instrumentation}
The observations were performed at the G1 observing pavilion with a 0.6m, f/12.5 Zeiss Cassegrain telescope. The Fiber Injection and Guiding Unit (hereafter FIGU) of the spectrograph is mounted in the Cassegrain focus of the telescope. The FIGU is connected to the calibration unit (ThAr hollow cathode lamp, tungsten lamp, blue LED) in the control room and to the echelle spectrograph itself in the cellar below the dome. The beam entering the spectrograph from the fiber is collimated by an achromatic 125mm doublet. The collimated beam is dispersed by a high-efficiency R2 echelle grating with 79 grooves/mm. The echelle spectrum is then cross-dispersed by a prism and imaged by a f/1.8 Canon lens on a CCD chip (Thizy \& Cochard 2011). The CCD camera (ATIK 460EX) uses 2184$\times$1472 chip with 4.54 $\mu$m square pixels. 2$\times$2 binning is normally used. The CCD camera quantum efficiency has maximum above 75\% between 4900-5700 \AA. The CCD chip fully records blue echelle orders, while from about 5000 \AA~ the orders start to be vignetted. For the red-most order used, 7350-7600 \AA, the recorded spectrum is cut at about 20\% of the peak blaze intensity. The stars are guided using a sensitive video camera WATEC 120N (720$\times$480 pixel chip) attached to the FIGU.

The maximum resolution of the spectrograph reaches $R \sim$ 12000 around 4700 \AA~ and 6200 \AA~. The useful spectral range is mostly limited by the chromatic aberration of the Canon photolens. A new focal reducer was installed in the beginning of July 2015 increasing the instrument throughput by about 50\% compared to values given in Pribulla et al. (2015). 

The total throughput of the telescope -- spectrograph system was measured on several nights with good ($<2$ arcsec) seeing. The resulting values are 0.8-1.4\% at 4400 \AA~ ($B$ filter), and 1.8-3.3\% at 5500 \AA~ ($V$ filter). The RV system stability was checked by looking at zero point shifts when re-identifying and solving individual ThAr spectra. The rms of zero shifts after removing a linear night trend varied from 35 to 49 m.s$^{-1}$ on six nights in February and March 2015.


\begin{table}[t!]
\centering
\caption{The journal of RV observations of HAT-P-2. The RV data, published previously by Cs\'ak et al. (2014), which were used for comparison, are denoted by a \dag. The table gives heliocentric Julian dates (HJD), barycentric RVs,
and 1$\sigma$ uncertainties of the RV values.}
\label{HAT-RV}
\begin{tabular}{lll}
\hline
\hline
HJD		& RV			& $\pm1 \sigma$ \\
$-2~400~000$   	& [km.s$^{-1}$]		& [km.s$^{-1}$] \\
\hline
\hline
   56731.5593   &    -21.735   		&    0.238      \\
   56815.4019   &    -20.262   		&    0.164      \\
   56842.4179   &    -19.981   		&    0.165      \\
   57071.5495   &    -19.852   		&    0.192      \\
   57072.5885   &    -19.856   		&    0.225      \\
   57088.5956   &    -19.951   		&    0.168      \\
   57098.6305   &    -20.286   		&    0.161      \\
   57136.5713   &    -20.580   		&    0.171      \\
   57154.5015   &    -20.985   		&    0.135      \\
   57183.4395   &    -19.685   		&    0.220      \\
   57191.4186   &    -19.805   		&    0.154      \\
   57206.3939   &    -19.808   		&    0.139      \\
   57208.3729   &    -19.423   		&    0.151      \\
   57210.4086   &    -21.449   		&    0.176      \\
   57211.4156   &    -20.033   		&    0.207      \\
   57241.3296   &    -19.550   		&    0.186      \\
   57242.3416   &    -19.688   		&    0.148      \\
   57243.3275   &    -20.043   		&    0.149      \\
   57243.3620   &    -20.364   		&    0.151      \\
   57406.6677   &    -20.463            &    0.141      \\
   55979.6629   &    -19.294            &    0.549\dag  \\  
   55980.5735   &    -19.511            &    0.442\dag  \\      
   55987.6470   &    -20.900            &    0.183\dag  \\      
   55992.5750   &    -19.569            &    0.268\dag  \\      
   55993.6021   &    -21.102            &    0.170\dag  \\      
   55995.6301   &    -19.336            &    0.072\dag  \\      
   55996.6278   &    -19.265            &    0.144\dag  \\      
   55997.6450   &    -19.648            &    0.119\dag  \\      
   55998.6025   &    -19.955            &    0.125\dag  \\      
   55999.6776   &    -20.441            &    0.135\dag  \\      
   56000.5263   &    -19.527            &    0.107\dag  \\      
   56008.6151   &    -19.270            &    0.184\dag  \\      
   56009.6110   &    -19.993            &    0.079\dag  \\      
   56012.5604   &    -19.447            &    0.166\dag  \\      
   56015.5193   &    -20.051            &    0.104\dag  \\      
   56048.4891   &    -19.779            &    0.102\dag  \\      
   56049.4024   &    -20.352            &    0.050\dag  \\      
   56053.4139   &    -19.231            &    0.228\dag  \\      
   56059.4339   &    -19.537            &    0.152\dag  \\      
\hline          
\end{tabular}
\end{table} 

\begin{table}[t!]
\centering
\caption{The journal of RV observations of WASP-14. The RV data, published previously by Joshi et al. (2009), which were used for comparison, are denoted by a \dag. The table gives heliocentric Julian dates (HJD), barycentric RVs and 1$\sigma$ uncertainties of the RV values.}
\label{WASP-RV}
\begin{tabular}{lll}
\hline
\hline
HJD		& RV			& $\pm1 \sigma$\\
$-2~400~000$   	& [km.s$^{-1}$]		& [km.s$^{-1}$]\\
\hline
\hline
   57070.6324   &    -4.322             &    0.231     \\
   57071.5139   &    -6.022             &    0.200     \\
   57072.5462   &    -4.430             &    0.286     \\
   57088.5566   &    -4.349             &    0.204     \\
   57098.5739   &    -5.709             &    0.186     \\
   57101.5074   &    -3.887             &    0.181     \\
   57102.5292   &    -5.859             &    0.191     \\
   57123.5363   &    -4.587             &    0.271     \\
   57136.4623   &    -5.810             &    0.183     \\
   57137.4698   &    -4.061             &    0.202     \\
   57138.4578   &    -5.856             &    0.183     \\
   57145.4533   &    -5.815             &    0.195     \\
   57154.4538   &    -6.094             &    0.169     \\
   57183.3532   &    -5.272             &    0.449     \\
   57191.3552   &    -3.867             &    0.226     \\
   57206.3465   &    -5.872             &    0.204     \\
   57207.3668   &    -4.066             &    0.169     \\
   57209.3837   &    -3.513             &    0.222     \\
   57210.3703   &    -5.476             &    0.251     \\  
   57406.6284   &    -4.175             &    0.164     \\ 
   54461.7400   &    -5.870             &    0.008\dag \\    
   54462.7710   &    -4.118             &    0.005\dag \\    
   54465.7490   &    -4.848             &    0.007\dag \\    
   54466.7850   &    -5.403             &    0.004\dag \\    
   54490.7700   &    -5.620             &    0.008\dag \\     
   54490.7820   &    -5.638             &    0.008\dag \\    
   54508.5938   &    -5.356             &    0.011\dag \\    
   54509.5238   &    -5.065             &    0.011\dag \\    
   54509.5830   &    -4.831             &    0.009\dag \\    
   54510.5074   &    -4.587             &    0.012\dag \\    
   54510.5599   &    -4.720             &    0.010\dag \\    
   54510.6509   &    -4.879             &    0.009\dag \\    
   54510.6569   &    -4.895             &    0.010\dag \\    
   54510.6630   &    -4.913             &    0.011\dag \\    
   54510.6860   &    -4.975             &    0.010\dag \\    
   54510.7215   &    -5.090             &    0.010\dag \\    
   54510.7274   &    -5.108             &    0.011\dag \\    
   54511.5572   &    -5.686             &    0.010\dag \\    
   54511.6822   &    -5.342             &    0.011\dag \\    
   54512.5406   &    -4.217             &    0.010\dag \\    
   54512.5682   &    -4.269             &    0.012\dag \\    
   54512.6822   &    -4.439             &    0.010\dag \\    
   54512.7152   &    -4.508             &    0.010\dag \\    
   54515.6212   &    -5.895             &    0.009\dag \\    
   54518.6525   &    -4.585             &    0.011\dag \\    
   54524.6750   &    -5.975             &    0.010\dag \\    
   54525.6755   &    -4.056             &    0.009\dag \\    
\hline                                                 
\end{tabular}
\end{table} 

\begin{table}[t!]
\centering
\caption{The journal of RV observations of XO-3. The RV data, published previously by H\'ebrard et al. (2008), which were used for comparison, are denoted by a \dag. The table gives heliocentric Julian dates (HJD), barycentric RVs and 1$\sigma$ uncertainties of the RV values.}
\label{XO-RV}
\begin{tabular}{lll}
\hline
\hline
HJD		& RV			& $\pm1 \sigma$ \\
$-2~400~000$   	& [km.s$^{-1}$]		& [km.s$^{-1}$] \\
\hline
\hline
   57061.2833   &    -12.244            &    0.657      \\     
   57068.3122   &    -12.943            &    0.263      \\     
   57069.3676   &    -09.882            &    0.242      \\     
   57088.2565   &    -10.775            &    0.280      \\     
   57098.2888   &    -10.112            &    0.215      \\     
   57100.2776   &    -12.334            &    0.276      \\     
   57102.3268   &    -12.190            &    0.253      \\     
   57105.3041   &    -11.534            &    0.214      \\     
   57241.5600   &    -10.548            &    0.168      \\     
   57242.5539   &    -11.571            &    0.167      \\     
   57243.5650   &    -12.272            &    0.209      \\     
   57327.5442   &    -11.471            &    0.159      \\     
   57328.2816   &    -09.731            &    0.403      \\     
   57329.3166   &    -12.725            &    0.181      \\     
   57360.4247   &    -10.735            &    0.166      \\     
   57371.6476   &    -12.373            &    0.303      \\     
   57373.3016   &    -11.416            &    0.170      \\     
   57380.2085   &    -11.905            &    0.358      \\     
   57389.3851   &    -11.846            &    0.205      \\     
   57406.2062   &    -12.919            &    0.264      \\     
   54496.2649   &    -12.723            &    0.050\dag  \\       
   54497.2609   &    -10.156            &    0.029\dag  \\       
   54499.2765   &    -13.006            &    0.030\dag  \\       
   54501.2926   &    -12.433            &    0.031\dag  \\       
   54501.4628   &    -12.756            &    0.033\dag  \\       
   54502.2730   &    -13.068            &    0.024\dag  \\       
   54503.2614   &    -10.936            &    0.030\dag  \\       
   54503.4700   &    -10.182            &    0.040\dag  \\       
   54504.4321   &    -12.398            &    0.028\dag  \\       
   54505.2889   &    -13.132            &    0.023\dag  \\       
   54506.2904   &    -11.593            &    0.025\dag  \\       
   54511.4534   &    -13.041            &    0.038\dag  \\       
   54512.4618   &    -12.246            &    0.036\dag  \\       
   54513.3091   &    -10.360            &    0.052\dag  \\       
   54516.3517   &    -10.176            &    0.046\dag  \\       
   54516.4540   &    -10.267            &    0.071\dag  \\       
   54551.3044   &    -10.316            &    0.032\dag  \\       
   54553.3002   &    -13.135            &    0.033\dag  \\       
   54554.3114   &    -11.004            &    0.019\dag  \\       
\hline
\end{tabular}
\end{table} 

\section{Observations and data analysis}
\label{obs}
Using our instrumentation we obtained 20 RV measurements per planetary system. The journals of RV observations of the HAT-P-2, WASP-14 and XO-3 systems are summarized in Tables \ref{HAT-RV}, \ref{WASP-RV} and \ref{XO-RV}, respectively\footnote{HJD values in this paper are in the UTC system.}. The observations were obtained between March 14, 2014 and January 18, 2016 for HAT-P-2, February 16, 2015 and January 18, 2016 for WASP-14 and between February 7, 2015 and January 18, 2016 for XO-3. The exposure times were 900 s in all cases. Usually three raw spectra per object were obtained during one observing night and subsequently combined via IRAF\footnote{see {\tt http://iraf.noao.edu/}} task {\tt{combine}} to increase the SNR. Three 900 s exposures correspond to about 0.5-1.5 \% of the orbital period.      

The data were reduced using IRAF package tasks, Linux shell scripts and FORTRAN programs as described in Pribulla et al. (2015). In the first step, master dark frames were produced. In the second step, the photometric calibration of the frames was done using dark and flat-field frames. Bad pixels were cleaned using a bad pixel mask, cosmic hits were removed using the program of Pych (2004). Photometrically calibrated frames were combined to increase the SNR. Order positions were defined by fitting Chebyshev polynomials to tungsten-lamp and blue LED spectrum. In the following step, scattered light was modeled and subtracted. Aperture spectra were then extracted for both object and ThAr frames and then the resulting 2D spectra were dispersion solved. 2D spectra were finally combined to 1D spectra rebinned to 4220 \AA~ to 7620 \AA~ wavelength range with a 0.1 \AA~ step ($\sim$ 3-6 times the spectral resolution). Spectra of all three targets were analyzed using the broadening function (hereafter BF) technique developed by Rucinski (1992). The BFs have been extracted in 4900 \AA~ - 5510 \AA~ spectral range (free of Balmer lines) with a velocity step of 5.8 km.s$^{-1}$ (matching pixel resolution of the spectra). 
Because the projected rotational velocity of all three systems is smaller than the spectral resolution the Gaussian profiles have been fitted to the extracted BFs in all cases to arrive at RV. 

Because the formal RV uncertainties found using the BF approach are hard to quantify and they depend on BF smoothing they were determined from SNR as follows. First, we determined SNR at 5200 \AA~ and calculated RV errors as 1/SNR (see Eq. \ref{hatzes} in Sect. \ref{cnc}). Subsequently, we fitted RV observations with initial uncertainties using the code JKTEBOP (described in the next paragraph) and from the best fit we obtained reduced $\chi^2$. In the next step, we rescaled all uncertainties to get $\chi^2 = 1$. SNR of the spectra can be easily obtained from the $1\sigma$ uncertainties as $C/\sigma$, where the scaling constant $C$ was found to be 10.59 km.s$^{-1}$ for HAT-P-2, 7.52 km.s$^{-1}$ for WASP-14 and 8.66 km.s$^{-1}$ for XO-3. The systematic uncertainties resulting from the limited stability of the RV system are significantly lower than random errors and were neglected in the analysis. We then fitted our RV data with correct uncertainties using the code JKTEBOP. 

In general, the JKTEBOP code\footnote{{\tt http://www.astro.keele.ac.uk/jkt/codes/jktebop.html}} (Southworth, Maxted \& Smalley 2004) is used to fit a model to the light curves of detached eclipsing binary stars in order to derive the radii of the stars as well as various other quantities. It is also excellent for transiting extrasolar planetary systems. JKTEBOP can fit RVs simultaneously with a light curve; hence the orbital parameters of the three transiting planets were calculated from the RV data, together with the photometric data. The transit light curve of HAT-P-2b was taken from the public archive of the HATNet project\footnote{see {\tt http://hatnet.org/planets/followup-data.html}}. The Sloan \textit{z}-band photometry (effective central wavelength 966.5 nm, FWHM 255.8 nm) was taken with the KeplerCam detector on the Fred L. Whipple Observatory (hereafter FLWO) 1.2m telescope on Mount Hopkins, Arizona (see e.g., Holman et al. 2007) on April 22, 2007. Three Bessell \textit{R}-band transit light curves of WASP-14b were adopted from Raetz et al. (2015). The light curves were taken with the STK camera (Mugrauer \& Berthold 2010) on the 0.6/0.9m telescope at the Gro{\ss}schwabhausen Observatory\footnote{{\tt http://www.astro.uni-jena.de/index.php/gsh-home.html}} in Germany (hereafter GSH) on March 11, 20 and 29, 2011. The Bessell \textit{V}-band transit light curve of XO-3b was observed at the Teide Observatory (Island of Tenerife, Canary Islands -- Spain) on November 17, 2015. The instrument was the OGS Spectrograph, installed on the 1.0m telescope (Koehler 1997), called Optical Ground Station\footnote{\tt{http://www.iac.es/eno.php?op1=3\&op2=6\&lang=en\&id=7}} (hereafter OGS), which was used in imaging mode. 

We fitted RV data from different sources, simultaneously with the exact same photometric data per object. First, we fitted only our G1 RV data, simultaneously with the photometric data. Except WASP-14b, 8 free parameters were adjusted during the fitting procedure: the sum of fractional radii ($(R_\mathrm{s}+R_\mathrm{p})/a$), the ratio of the radii ($R_\mathrm{p}/R_\mathrm{s}$), the orbit inclination angle ($i$), the orbit eccentricity ($e$), the longitude of the periastron ($\omega$), the systemic velocity ($\gamma$), the RV semi-amplitude of the parent star ($K$) and the light scale factor ($L_\mathrm{sf}$). In the case of WASP-14b, since its orbit eccentricity is relatively smaller, we replaced $e$ and $\omega$ with $k=e~\cos \omega$ and $h=e~\sin \omega$, and since we had three transit light curves, we adjusted the orbital period ($P_\mathrm{orb}$) as well. Initial parameters for the fitting procedure were taken from published papers (Bakos et al. 2007; P\'al et al. 2010; Lewis et al. 2013; Joshi et al. 2009; Husnoo et al. 2011; Wong et al. 2014, 2015; Winn et al. 2008, 2009 and H\'ebrard et al. 2008). The quadratic limb-darkening (hereafter LD) law was assumed for the parent stars, which is a better choice to represent the observations than a simple linear law (see e.g., Raetz et al. 2014). The LD coefficients were fixed during the fitting procedure and their values were calculated for the $z$-band, $R$-band, or $V$-band using the on-line applet EXOFAST-limbdark\footnote{{\tt http://astroutils.astronomy.ohio-state.edu/exofast/}} (Eastman, Gaudi \& Agol 2013). The software interpolates tables published by Claret \& Bloemen (2011) based on the stellar parameters Fe/H (M/H), $T_\mathrm{eff}$ and $\log g$. To estimate the uncertainties in fitted parameters we used JKTEBOP-task No. 8. This finds the best fit and then uses Monte Carlo simulations to estimate the uncertainties in the parameters. The best-fit model is re-evaluated at the phases of the actual observations. Gaussian simulated observational noise is added and the result refitted. This process is repeated and the range in parameter values found gives the uncertainty in that parameter. We first applied a sequence of iterations. At the first sequence 1000 iteration steps were used and each next sequence contained 1000 more iterations. After 8000 iterations the uncertainties in the parameters varied only slightly. To obtain final results we applied 10~000 iteration steps.  

Subsequently, we fitted previously published RV data from other sub-meter-, meter- and two-meter-class telescopes, simultaneously with the photometric data, in the same way as we described above, in order to compare the results. In the case of HAT-P-2 we used the RV data published by Cs\'ak et al. (2014). Observations were carried out between September 2011 and April 2013 at two locations in Hungary: at the Gothard Astrophysical Observatory, Szombathely (hereafter GAO) and at the Piszk\'estet\H{o} Mountain Station of Hungarian Academy of Sciences (hereafter PO). GAO has a newly installed 0.5m diameter f/9 telescope for spectroscopic observations. At PO a 1m telescope was used. The spectrograph was the same fiber-fed instrument at both locations, the eShel system of the French Shelyak Instruments, however, the spectrum was recorded by a QSI 532ws CCD camera (Kodak KAF1600 CCD chip). For WASP-14 we adopted the RV data published by Joshi et al. (2009), which were collected with the FIbre-fed Echelle Spectrograph (hereafter FIES) in medium-resolution mode\footnote{FIES is a cross-dispersed high-resolution fibre-fed echelle spectrograph. The currently installed fibers offer spectral resolution of $R=67~000$ in high-resolution mode, $R=46~000$ in medium-resolution mode and $R=25~000$ in low-resolution mode. The entire spectral range 3700-7300 \AA~ is covered without gaps.} (Telting et al. 2014), mounted on the 2.5m Nordic Optical Telescope (hereafter NOT) and with the SOPHIE echelle spectrograph in high-efficiency mode\footnote{SOPHIE is a cross-dispersed fiber-fed echelle spectrograph, which can operate in two observation modes: in high-efficiency mode ($R=39~000$) to favor high throughput, or in high-resolution mode ($R=75~000$) to favor better precision. SOPHIE covers the spectral range 3870-6940 \AA.} (Perruchot et al. 2008) on the 1.93m telescope at the Haute-Provence Observatory (hereafter OHP), during February 12 and 29, 2008. In the case of XO-3 we used the RV data published by H\'ebrard et al. (2008). Observations were acquired during February and March, 2008 at the same observatory and telescope with the SOPHIE echelle spectrograph in high-resolution mode.          

Finally, we combined and analyzed all used data per object. RV offsets among different telescopes were corrected adjusting the systemic RV parameter only. Up to five iteration steps were enough for our purposes.                            


\section{Results}

\subsection{HAT-P-2b}
\label{SECT-HAT}
\begin{figure}[t!]
\centering
\includegraphics[width=80mm]{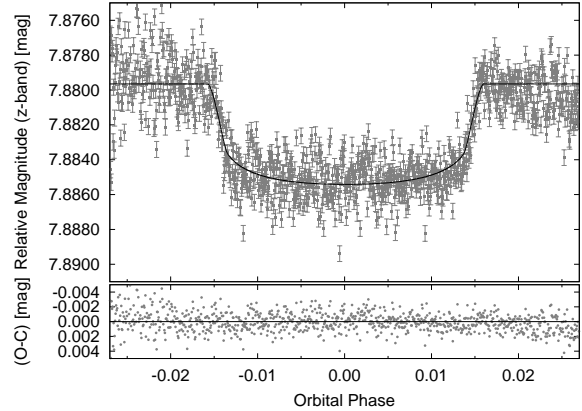}   
\caption{The phase-folded $z$-band FLWO light curve of HAT-P-2b, taken from the public archive of the HATNet project, showing the transit event. The observed light curve is overplotted with our best JKTEBOP fit model. Corresponding residuals are also shown (bottom). The model was calculated based on simultaneous fit to the photometric and all RV data (see Section \ref{SECT-HAT} and Table \ref{HAT-RV}).}
\label{HAT-LC}
\end{figure}

HAT-P-2b is a massive ($M_\mathrm{p} \simeq 9.09 M_\mathrm{J}$; P\'al et al. 2010) transiting extrasolar planet on an eccentric orbit ($e \simeq 0.50910$; Lewis et al. 2013). It was discovered by Bakos et al. (2007). The planet orbits the host star ($V=9.69$ mag, $B=9.15$ mag, F8V, $v\sin i \simeq 20.8$ km.s$^{-1}$) with a period of about 5.633 days causing transits with a depth of 5 mmag and duration of about 4.25 hours. The planetary transit was followed up with the 1.2m telescope and its KeplerCam detector at FLWO in the Sloan $z$-band (in Figure \ref{HAT-LC} the transit light curve is phased with the orbital period of $P_\mathrm{orb}=5.633472$ days and $T_0=2454387.4937$ HJD). The secondary eclipse and phase curve observations of HAT-P-2b were obtained in the 3.6, 4.5, 5.8 and 8.0 $\mu$m bands of the Spitzer Space Telescope (Lewis et al. 2013). The planetary nature of the transiting object was confirmed by RV measurements with the HIRES instrument (Vogt et al. 1994) on the W.M. Keck telescope and with the Hamilton echelle spectrograph at the Lick Observatory (Vogt 1987). 

HAT-P-2b is an interesting planet, since the high eccentricity at the relatively small orbital distance ($a \simeq 0.06878$ AU; P\'al et al. 2010), raising the possibility that it migrated through planet-planet scattering or Kozai oscillations accompanied by tidal dissipation. Either of mechanisms can be investigated by exploiting the Rossiter-McLaughin (RM) effect, which is an "anomalous Doppler shift" during the transit (see e.g., Gaudi \& Winn 2007). Planet-planet scattering (see e.g., Chatterjee et al. 2008) or Kozai mechanism (see e.g., Wu, Murray \& Ramsahai 2007) can significantly tilt the orbit plane of the planet relative to the equatorial plane of the host star and this feature is well seen on the shape of the RM effect. The spin-orbit misalignment of the HAT-P-2 system was investigated by Winn et al. (2007) and Loeillet et al. (2008), however, both papers reported an angle consistent with zero. Lewis et al. (2013) reported an evidence for a long term linear trend in the radial velocity data, which suggests the presence of another substellar companion in the HAT-P-2 planetary system. It could have caused migration of HAT-P-2b via Kozai oscillations.

\begin{table}[b!]
\centering
\caption{Physical and orbital parameters of the HAT-P-2 system summarized from papers Bakos et al. (2007) [B07], P\'al et al. (2010) [P10] and Lewis et al. (2013) [L13].}
\label{HAT-DATA}
\begin{tabular}{lll}
\hline
\hline
Parameter					& Value			& Ref      \\
\hline
\hline
Zero transit time $T_0$ [red. HJD]$^{a}$ 	& 54387.4937(7)		& [P10]    \\
Orbital period $P_\mathrm{orb}$ [d]		& 5.633472(6)		& [P10]    \\
Normalized semimajor axis $a/R_\mathrm{s}$	& 8.9(3)		& [P10]	   \\
Ratio of the radii $R_\mathrm{p}/R_\mathrm{s}$	& 0.0722(6)		& [P10]    \\
Inclination $i$ [deg] 	 			& 86.7(8)		& [P10]    \\
Eccentricity $e$				& 0.50910(48)		& [L13]    \\
Periastron longitude $\omega$ [deg]		& 188.09(39)		& [L13]    \\
Systemic velocity $\gamma$ [m.s$^{-1}$]		& -278(20)		& [B07]    \\
RV semi-amplitude $K$ [m.s$^{-1}$]		& 983(17)		& [P10]    \\   
Metallicity (Fe/H)				& +0.12(8)		& [B07]    \\
Effective temperature $T_\mathrm{eff}$ [K]	& 6290(110)		& [B07]    \\
Star surface gravity $\log g$			& 4.22(14)		& [B07]    \\ 
\hline          
\end{tabular}
\raggedright{\scriptsize{$^{a}$red. HJD = HJD $-$ 2~400~000}}
\end{table} 

\begin{figure*}[t!]
\centering
\centerline{
\includegraphics[width=57mm,height=57mm]{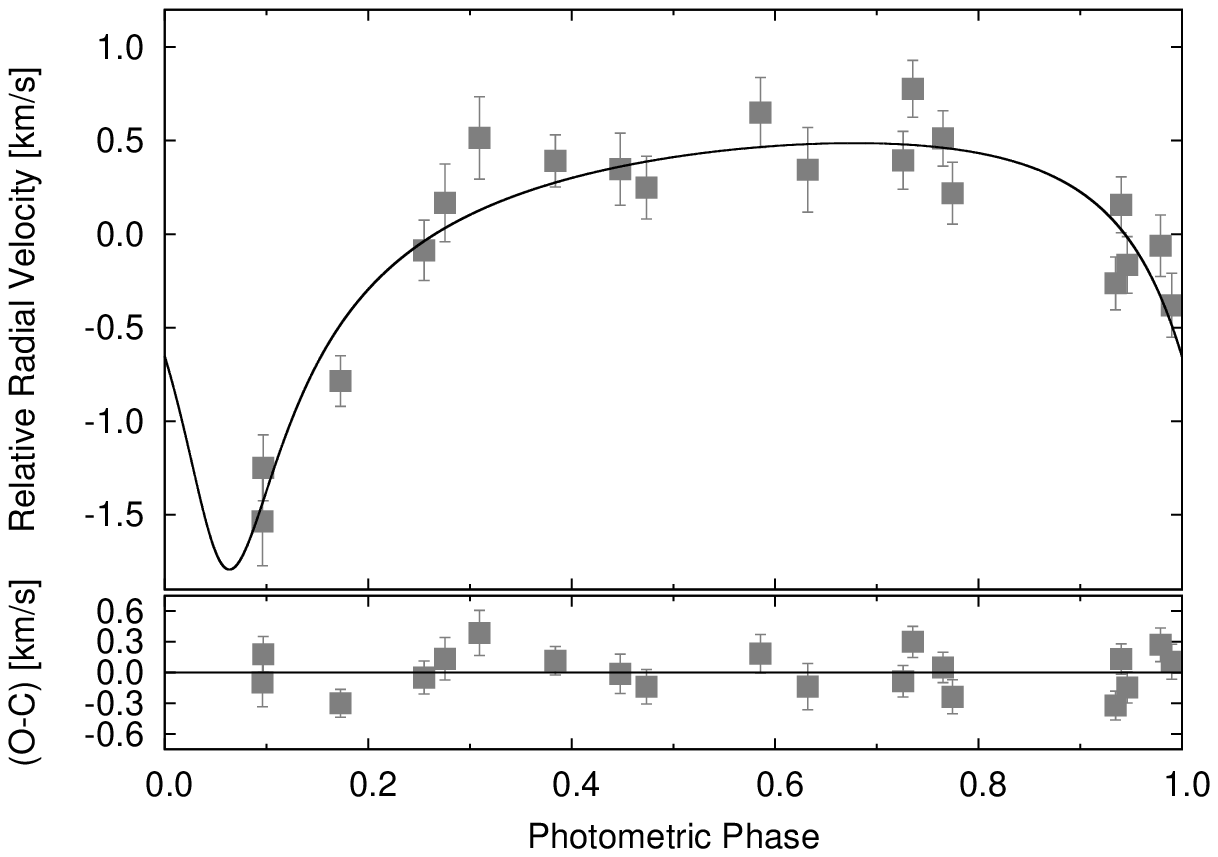}
\includegraphics[width=57mm,height=57mm]{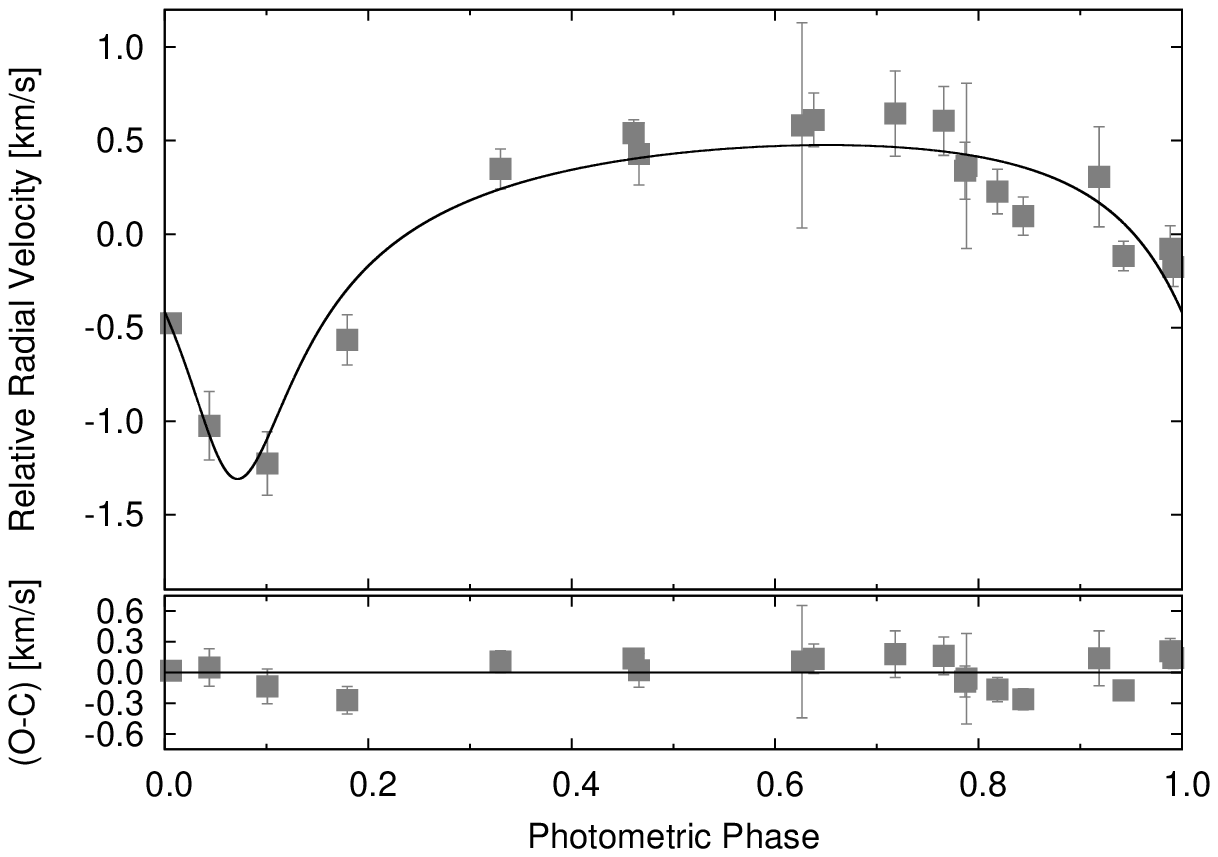}
\includegraphics[width=57mm,height=57mm]{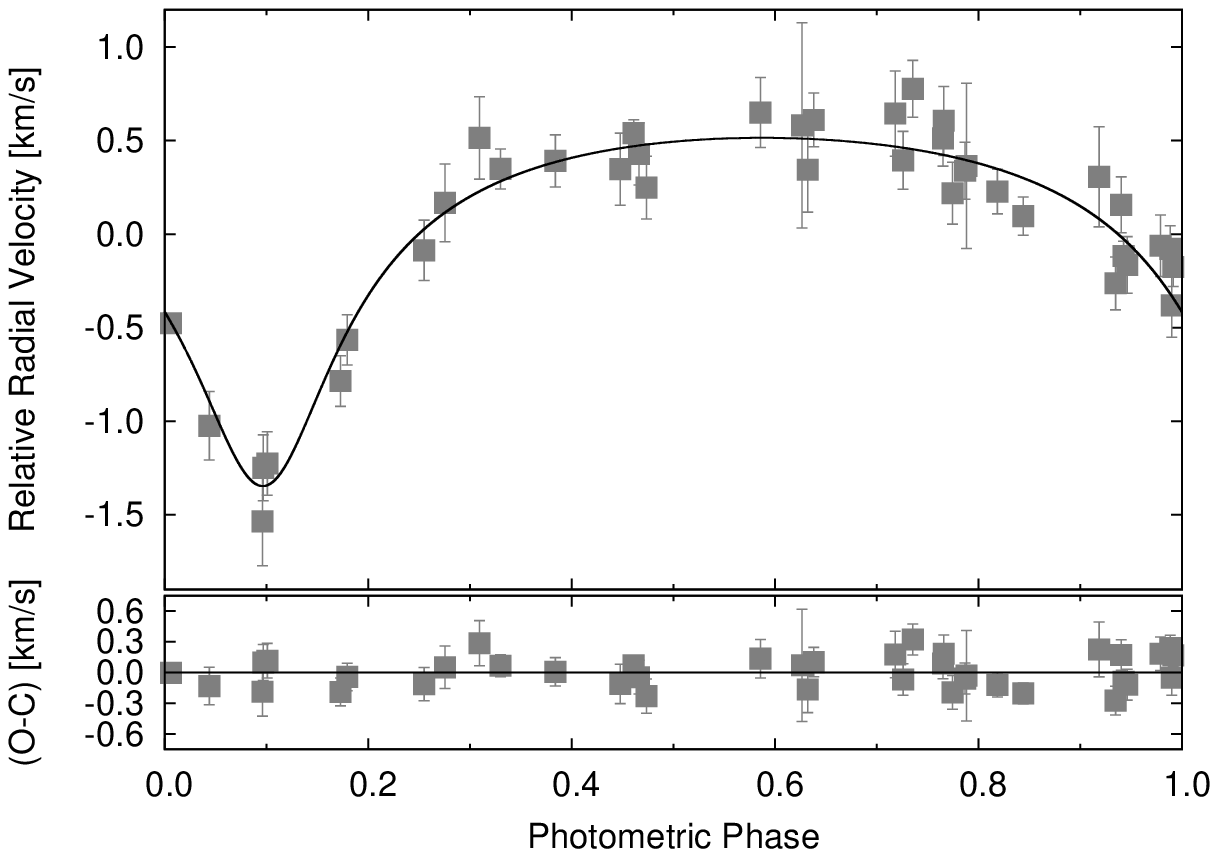}
}   
\caption{RV measurements of HAT-P-2 phased with the orbital period of $P_\mathrm{orb}=5.633472$ days. The zero point in phase corresponds to the epoch of midtransit $T_0=2454387.4937$ HJD. The RV measurements are overplotted with our best JKTEBOP fit model. Corresponding residuals are also shown (bottom). The model was calculated based on simultaneous fit to the photometric data and RV data obtained at G1 (left panel), at GAO and PO (middle panel) and at the all of here mentioned observatories (right panel).}
\label{HAT-RVFIT}
\end{figure*}

The semi-amplitude of the RV is $K=983(17)$ m.s$^{-1}$ (P\'al et al. 2010), therefore it is an easily measurable object for our echelle spectrograph, mounted on the 0.6m telescope (Pribulla et al. 2015). Together we worked with 39 (20+19) RV measurements (see Table \ref{HAT-RV}). The resulting G1 RVs show an averaged scatter of about 170 m.s$^{-1}$. First, we fitted only our G1 RV data, simultaneously with the photometric data (Figure \ref{HAT-LC}). Initial parameters for the JKTEBOP fitting were taken from published papers by Bakos et al. (2007), P\'al et al. (2010) and Lewis et al. (2013). These are summarized in Table \ref{HAT-DATA}. The sum of fractional radii ($(R_\mathrm{s}+R_\mathrm{p})/a$) was calculated from the normalized semi-major axis ($a/R_\mathrm{s}$) and from the ratio of the radii ($R_\mathrm{p}/R_\mathrm{s}$). The interpolated quadratic LD coefficients for the $z$-band were 0.186 and 0.302. The fitted RV data are shown in Figure \ref{HAT-RVFIT} -- left panel and the best fit parameters are summarized in Table \ref{ALL-DATA}. Subsequently, we fitted GAO and PO RV data, simultaneously with the photometric data. The fitted RV data are shown in Figure \ref{HAT-RVFIT} -- middle panel and the best fit parameters are summarized in Table \ref{ALL-DATA}. The averaged RV scatter is about 177 m.s$^{-1}$, comparable with our G1 RV scatter. We note, however, that the accuracy of these data is sometimes better than 100 m.s$^{-1}$, while in the case of the data from G1 the accuracy is sometimes better than 150 m.s$^{-1}$ only. The best accuracy in the case of the data from G1 is 135 m.s$^{-1}$, while for the data from GAO and PO it is 50 m.s$^{-1}$. Finally, we simultaneously fitted photometric and all RV data. The fitted RV data are shown in Figure \ref{HAT-RVFIT} -- right panel and the best fit parameters are summarized in Table \ref{ALL-DATA}.               

In Figure \ref{HAT-RVFIT} we can see that JKTEBOP fit models are similar. Table \ref{ALL-DATA} shows that the parameter values derived from the G1 RV data are, in general, consistent with the values adopted from literature (Bakos et al. 2007; P\'al et al. 2010; Lewis et al. 2013). The best consistency is seen in the case of the systemic velocity $\gamma$ ($-278(20)$ versus $-279(44)$ ~[m.s$^{-1}$]), or in the case of the RV semi-amplitude $K$ ($983(17)$ versus $1131(179)$ ~[m.s$^{-1}$]). On the other hand, the values of the orbit eccentricity $e$ ($0.50910(48)$ versus $0.58(4)$) are more inconsistent and we did not confirm the periastron longitude $\omega$ ($188.09(39)$ versus $165(6)$ ~[deg]; $\sim 3.8 \sigma$) and the ratio of the radii $R_\mathrm{p}/R_\mathrm{s}$ ($0.0722(6)$ versus $0.0677(2)$; $\sim 7.5 \sigma$). In comparison between the G1 and GAO/PO RV data, we can state that parameter values are determined similarly. This is mainly true for values of the orbit eccentricity $e$ ($0.58(4)$ versus $0.56(2)$) and for values of the periastron longitude $\omega$ ($165(6)$ versus $175(5)$ ~[deg]). The greatest difference is seen in values of the RV semi-amplitude $K$ ($1131(179)$ versus $904(52)$ ~[m.s$^{-1}$]). Scatter of the RV data points from the fit are, however, very similar (up to $\pm$500 m.s$^{-1}$; see Figure \ref{HAT-RVFIT}). 

\subsection{WASP-14b}
\label{SECT-WASP}
\begin{figure}[t!]
\centering
\includegraphics[width=80mm]{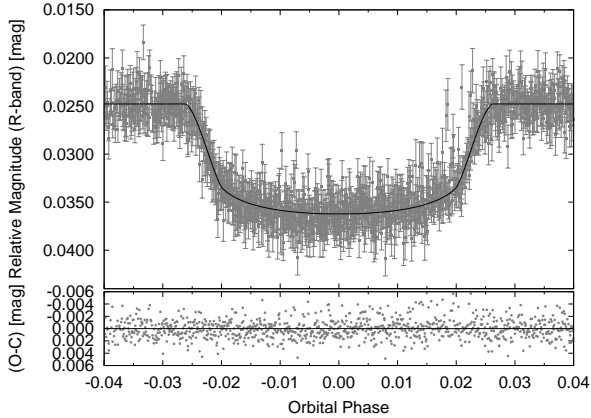}   
\caption{The phase-folded $R$-band STK light curve of WASP-14b, taken from the GSH archive, showing the transit event. The observed light curve is overplotted with our best JKTEBOP fit model. Corresponding residuals are also shown (bottom). The model was calculated based on simultaneous fit to the photometric and all RV data (see Section \ref{SECT-WASP} and Table \ref{WASP-RV}).}
\label{WASP-LC}
\end{figure}

The discovery of the 7.3 $M_\mathrm{J}$ exoplanet WASP-14b was reported by Joshi et al. (2009). The planet orbits the host star ($V=9.75$ mag, $B=10.19$ mag, F5V, $v\sin i \simeq 4.9$ km.s$^{-1}$) with a period of about 2.243 days and orbital eccentricity $e \simeq 0.08$ (Husnoo et al. 2011; Blecic et al. 2013; Wong et al. 2015). The orbiting planet causes transits with a depth of 11 mmag and duration of about 2.8 hours. Initial follow-up photometric observations of WASP-14b were carried out in the $V$ and $R$ passband by the robotic 2m Liverpool Telescope (see e.g., Gibson et al. 2008; Steele et al. 2008), equipped with the high-speed imaging RISE camera. Johnson et al. (2009) obtained the transit photometry with the University of Hawaii 2.2m telescope and the Orthogonal Parallel Transfer Imaging Camera (OPTIC; Tonry, Burke \& Schechter 1997) on Mauna Kea. Full-orbit phase curves of WASP-14b obtained by the Spitzer Space Telescope in the 3.6 and 4.5 $\mu$m bands were presented and analyzed by Wong et al. (2015). WASP-14b was also a subject of the framework of "Transit Timing Variations @ Young Exoplanet Transit Initiative" (TTV@YETI). 19 light curves of 13 individual transit events were collected during the period of seven years (2007 -- 2013). The transit ephemeris was refined, but significant evidence for TTV was not found (Raetz et al. 2015). For our purposes we adopted three \textit{R}-band transit light curves of WASP-14b from this study, taken on March 11, 20 and 29, 2011 with the STK camera on the 0.6/0.9m telescope at GSH. The phase-folded light curve, phased with the period of $P_\mathrm{orb}=2.243752$ days and $T_0=2454463.5758$ HJD is shown in Figure \ref{WASP-LC}. 

\begin{figure*}[t!]
\centering
\centerline{
\includegraphics[width=57mm,height=57mm]{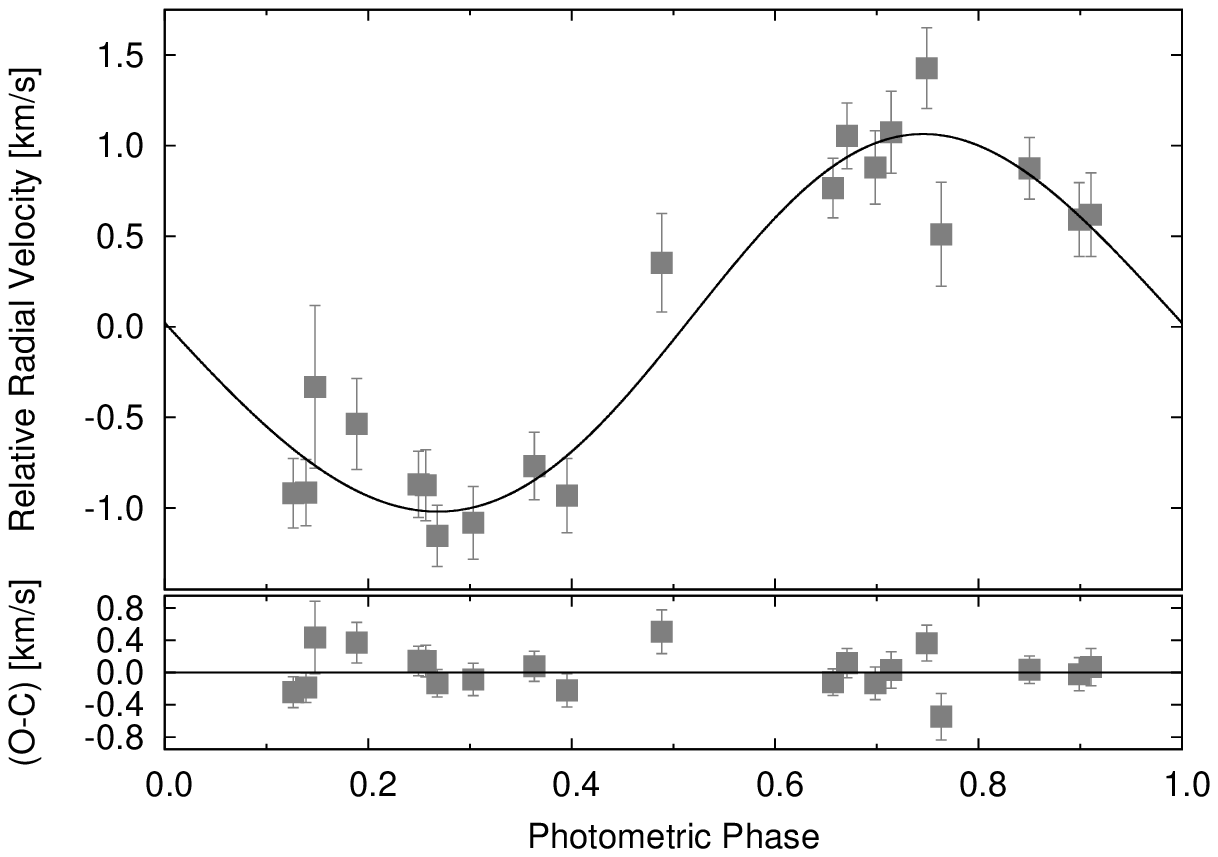}
\includegraphics[width=57mm,height=57mm]{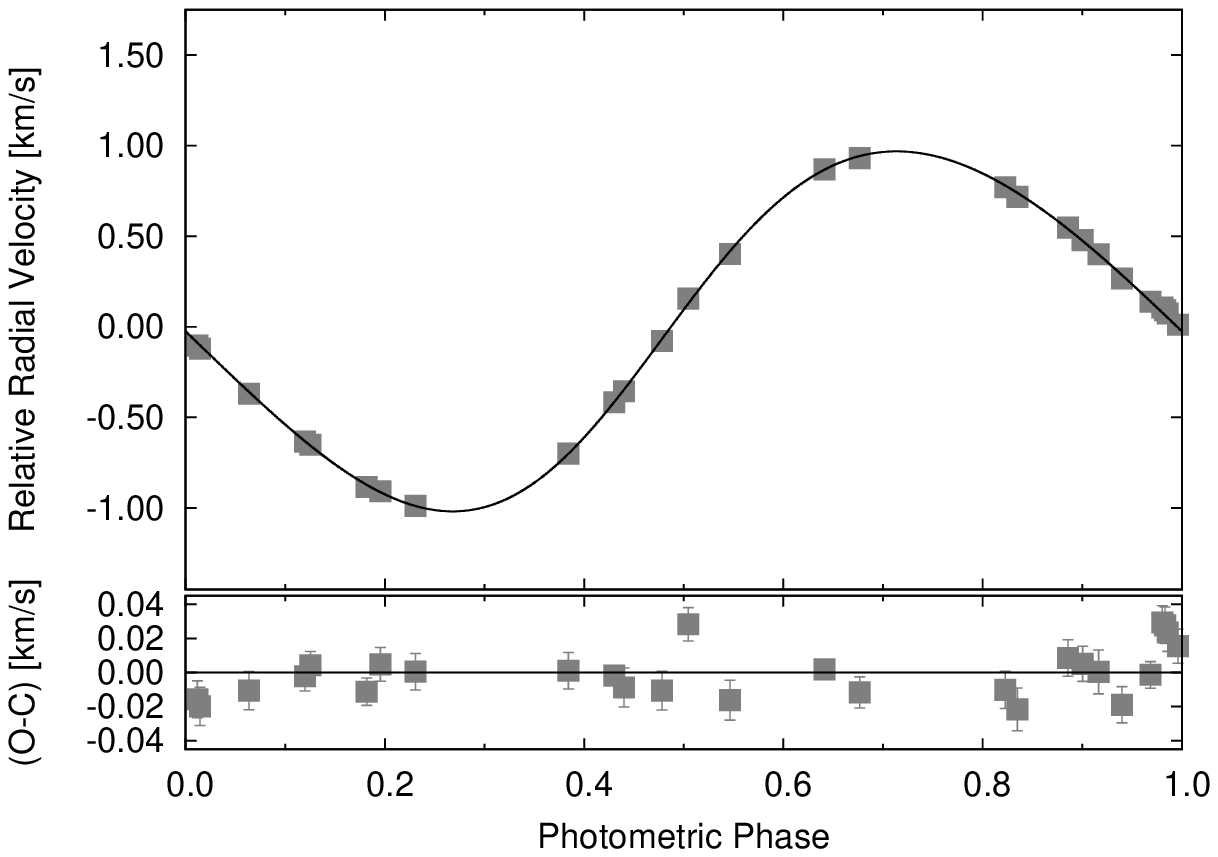}
\includegraphics[width=57mm,height=57mm]{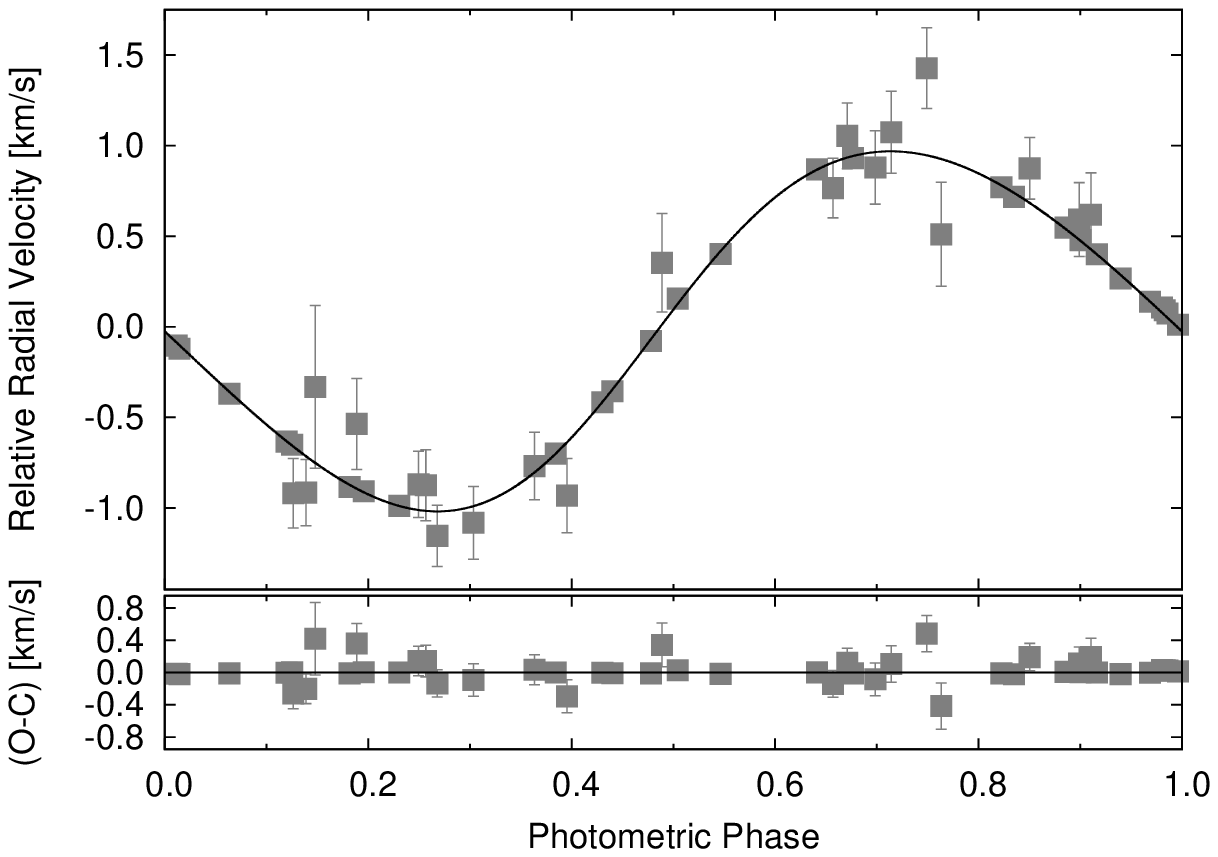}
}   
\caption{RV measurements of WASP-14 phased with the orbital period of $P_\mathrm{orb}=2.243752$ days. The zero point in phase corresponds to the epoch of midtransit $T_0=2454463.5758$ HJD. The RV measurements are overplotted with our best JKTEBOP fit model. Corresponding residuals are also shown (bottom). The model was calculated based on simultaneous fit to the photometric data and RV data obtained at G1 (left panel), at NOT and OHP (middle panel) and at the all of here mentioned observatories (right panel). Note that RV measurements obtained at NOT and OHP have too small uncertainties to discern.}
\label{WASP-RVFIT}
\end{figure*}

\begin{table}[b!]
\centering
\caption{Physical and orbital parameters of the WASP-14 system summarized from papers Joshi et al. (2009) [J09], Husnoo et al. (2011) [H11] and Wong et al. (2015) [W15].}
\label{WASP-DATA}
\begin{tabular}{lll}
\hline
\hline
Parameter						& Value			& Ref      \\
\hline
\hline
Zero transit time $T_0$ [red. HJD]$^{a}$		& 54463.5758(5)		& [J09]    \\
Orbital period $P_\mathrm{orb}$ [d]			& 2.243752(10)		& [J09]    \\
Orbital semimajor axis $a$ [AU]				& 0.0360(10)		& [J09]    \\
Stellar radius $R_\mathrm{s}$ [R$_{\odot}$]		& 1.30(6)		& [J09]    \\ 
Transit depth $(R_\mathrm{p}/R_\mathrm{s})^2$ [mag] 	& 0.0102(2)		& [J09]    \\
Inclination $i$ [deg] 	 				& 84.3(6)		& [J09]    \\
$k=e~\cos \omega$					& $-$0.02474(78)	& [W15]    \\
$h=e~\sin \omega$					& $-$0.0792(31)  	& [W15]    \\
Systemic velocity $\gamma$ [m.s$^{-1}$]			& $-$4985(3)		& [H11]    \\
RV semi-amplitude $K$ [m.s$^{-1}$]			& 991(3)		& [H11]    \\   
Metallicity (M/H)					& 0.0(2)		& [J09]    \\
Effective temperature $T_\mathrm{eff}$ [K]		& 6475(100)		& [J09]    \\
Star surface gravity $\log g$				& 4.0(2)		& [J09]    \\ 
\hline          
\end{tabular}
\raggedright{\scriptsize{$^{a}$red. HJD = HJD $-$ 2~400~000}}
\end{table}   

Spectroscopic follow-up observations started with measurements on 2.5m NOT (Joshi et al. 2009). A total of six spectra were obtained here with the FIES instrument. A further 21 RV points were obtained using the SOPHIE spectrograph on the 1.93m telescope at OHP. Similarly to HAT-P-2b, one of the interesting features of WASP-14b is the high orbital eccentricity at the relatively small orbital distance ($a \simeq 0.036$ AU; Joshi et al. 2009), raising the possibility the planet-planet scattering or Kozai oscillations. Based on the RM effect, Joshi et al. (2009) reported a measurement of the spin-orbit angle -14(21) [deg], which is too uncertain to draw any conclusions. Subsequently, Johnson et al. (2009) measured the RM effect using the High-Dispersion Spectrometer (see Noguchi et al. 2002) on the Subaru 8.2m telescope during the transit, and presented an evidence that the WASP-14 system has misaligned orbital and stellar rotational axes, with an angle about -33(7) [deg].                     
           
The semi-amplitude of the RV is about $K=991$ m.s$^{-1}$ (Husnoo et al. 2011; Blecic et al. 2013), which is similar to HAT-P-2. Together we worked with 47 (20+27) RV measurements (see Table \ref{WASP-RV}). Our RVs show an averaged scatter of about 220 m.s$^{-1}$. First, we fitted only our G1 RV data, simultaneously with the photometric data (Figure \ref{WASP-LC}). Initial parameters for the JKTEBOP fitting were taken from published papers by Joshi et al. (2009), Husnoo et al. (2011) and Wong et al. (2015). These are summarized in Table \ref{WASP-DATA}. The sum of fractional radii ($(R_\mathrm{s}+R_\mathrm{p})/a$) was calculated from the orbital semi-major axis $a$, the stellar radius $R_\mathrm{s}$ and from the ratio of the radii ($R_\mathrm{p}/R_\mathrm{s}$). The ratio of the radii was calculated from the transit depth $(R_\mathrm{p}/R_\mathrm{s})^2$. The interpolated quadratic LD coefficients for the $R$-band were 0.267 and 0.324. The fitted RV data are shown in Figure \ref{WASP-RVFIT} -- left panel and the best fit parameters are summarized in Table \ref{ALL-DATA}. At the next step, we fitted NOT and OHP RV data, simultaneously with the photometric data. The fitted RV data are shown in Figure \ref{WASP-RVFIT} -- middle panel and the best fit parameters are summarized in Table \ref{ALL-DATA}. In this case, the averaged RV scatter is only about 10 m.s$^{-1}$. The accuracy of the data from NOT and OHP is sometimes better than 8 m.s$^{-1}$ (4 m.s$^{-1}$ is the best accuracy), while in the case of the data from G1 the accuracy is sometimes better than 170 m.s$^{-1}$ (164 m.s$^{-1}$ is the best accuracy). Finally, we simultaneously fitted photometric and all RV data. The fitted RV data are shown in Figure \ref{WASP-RVFIT} -- right panel and the best fit parameters are summarized in Table \ref{ALL-DATA}.                 	  

Figure \ref{WASP-RVFIT} shows, however, that the JKTEBOP fit model from our RV values is similar to the JKTEBOP fit model from values adopted from Joshi et al. (2009). The resulting parameter values are very similar e.g., in the case of the systemic velocity $\gamma$ ($-4985(48)$ versus $-4985.0(1.7)$ ~[m.s$^{-1}$]), or in the case of the $k=e~\cos \omega$ ($-0.02(3)$ versus $-0.0254(9)$). The parameter $h=e~\sin \omega$ is more inconsistent ($-0.03(6)$ versus $-0.085(2)$). The value of the RV semi-amplitude $K$ derived from G1 RVs ($1041(54)$ ~[m.s$^{-1}$]), is far from the parameter $K$ derived from NOT/OHP RVs ($993(3)$ ~[m.s$^{-1}$]), however, if we consider $1\sigma$ error limits, these values are also in agreement. The situation is very similar, if we compare best fit parameter values derived from G1 RV data and literature parameter values. Furthermore, we can also easily see that the parameter values derived from G1 RVs are, in general, more uncertain. The scatter of the RV data points from the fit is also very different: $\pm$500 m.s$^{-1}$ in the case of G1 RVs and $\pm$40 m.s$^{-1}$ in the case of NOT/OHP RVs. 

\subsection{XO-3b}
\label{SECT-XO}
\begin{figure}[t!]
\centering
\includegraphics[width=80mm]{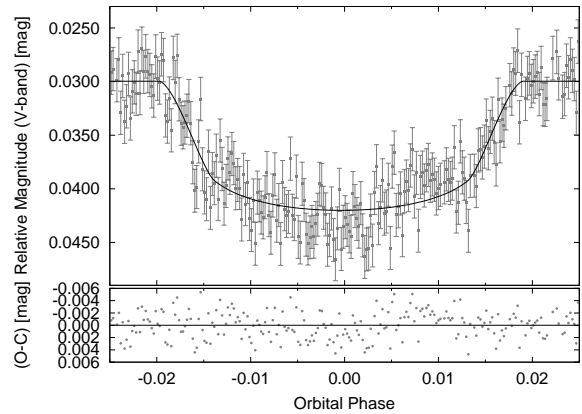}   
\caption{The phase-folded $V$-band OGS light curve of XO-3b, taken at the Teide Observatory, showing the transit event. The observed light curve is overplotted with our best JKTEBOP fit model. Corresponding residuals are also shown (bottom). The model was calculated based on simultaneous fit to the photometric and all RV data (see Section \ref{SECT-XO} and Table \ref{XO-RV}).}
\label{XO-LC}
\end{figure}

XO-3b is a massive ($M_\mathrm{p} \simeq 13.25 M_\mathrm{J}$; Johns-Krull et al. 2008) transiting planet, orbiting the parent star on an eccentric orbit ($e \simeq 0.2769$; Wong et al. 2014). It was discovered by Johns-Krull et al. (2008). The planet orbits the host star ($V=9.80$ mag, $B=10.25$ mag, F5V, $v\sin i \simeq 18.54$ km.s$^{-1}$) with a period of about 3.191 days causing transits with a depth of about 10 mmag and duration of about 2.86 hours. Since Johns-Krull et al. (2008) determined the planetary radius parameter $R_\mathrm{p}$ with a large uncertainty ($1.10-2.11$ $R_\mathrm{J}$), XO-3b was followed-up by Winn et al. (2008) with the 1.2m telescope and its KeplerCam detector at FLWO in the Sloan $z$-band, and with the 0.5m telescope, equipped with an SBIG ST-10 XME CCD camera (without filters) at the Wise Observatory (see Brosch et al. 2008) in Israel. Several additional photometric observations were also obtained at different observatories in the $I$- and $V$-band. Follow-up photometry obtained by Winn et al. (2008) strongly favor the planetary radius $R_\mathrm{p} \simeq 1.217$ $R_\mathrm{J}$. Wong et al. (2014) analyzed 12 secondary eclipse observations of XO-3b in the 4.5 $\mu$m Spitzer band and determined the upper limit of the periastron precession rate of $2.9 \times 10^{-3}$ deg per day. For our purposes we obtained a $V$-band transit light curve of XO-3b at the Teide Observatory on November 17, 2015 using the 1m OGS telescope. The phase-folded light curve, phased with the period of $P_\mathrm{orb}=3.191523$ days and $T_0=2454449.8681$ HJD is shown in Figure \ref{XO-LC}.   

\begin{figure*}[t!]
\centering
\centerline{
\includegraphics[width=57mm,height=57mm]{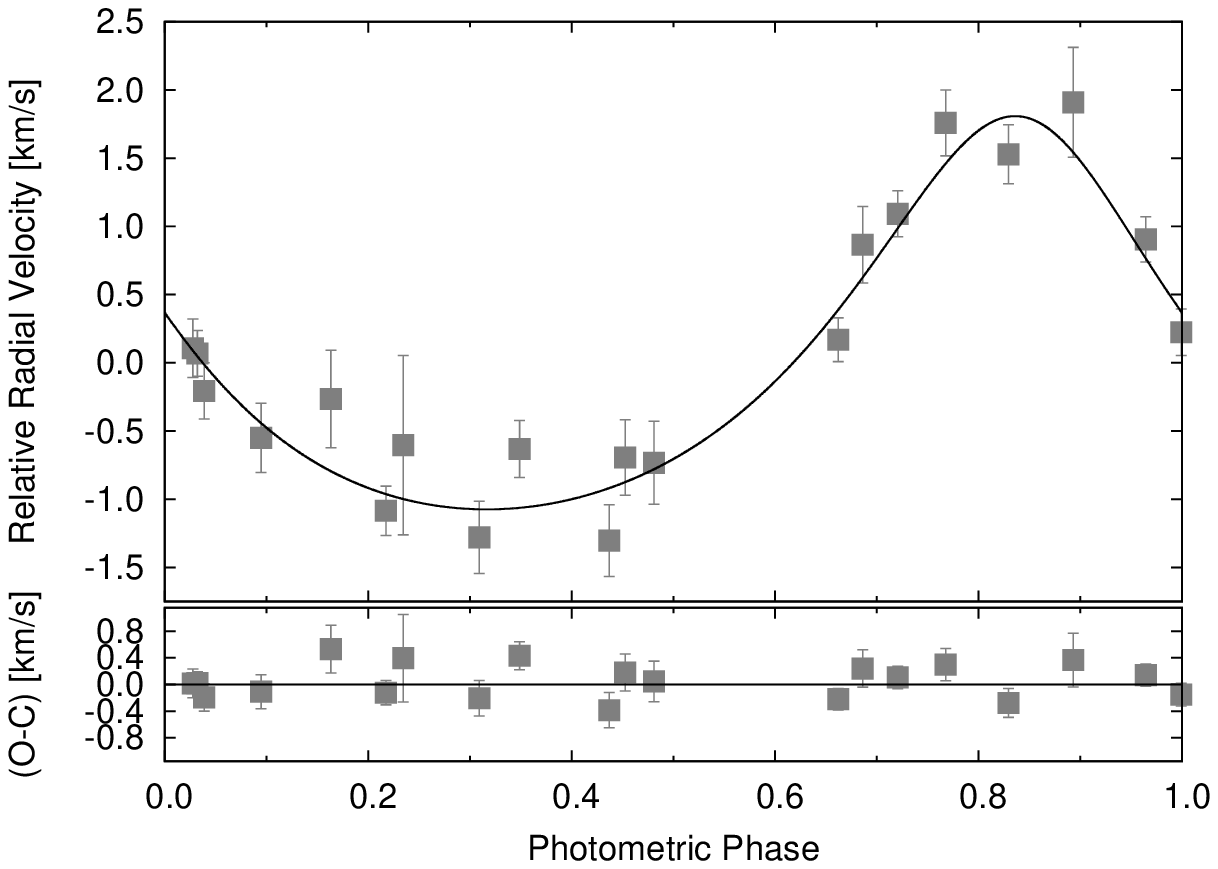}
\includegraphics[width=57mm,height=57mm]{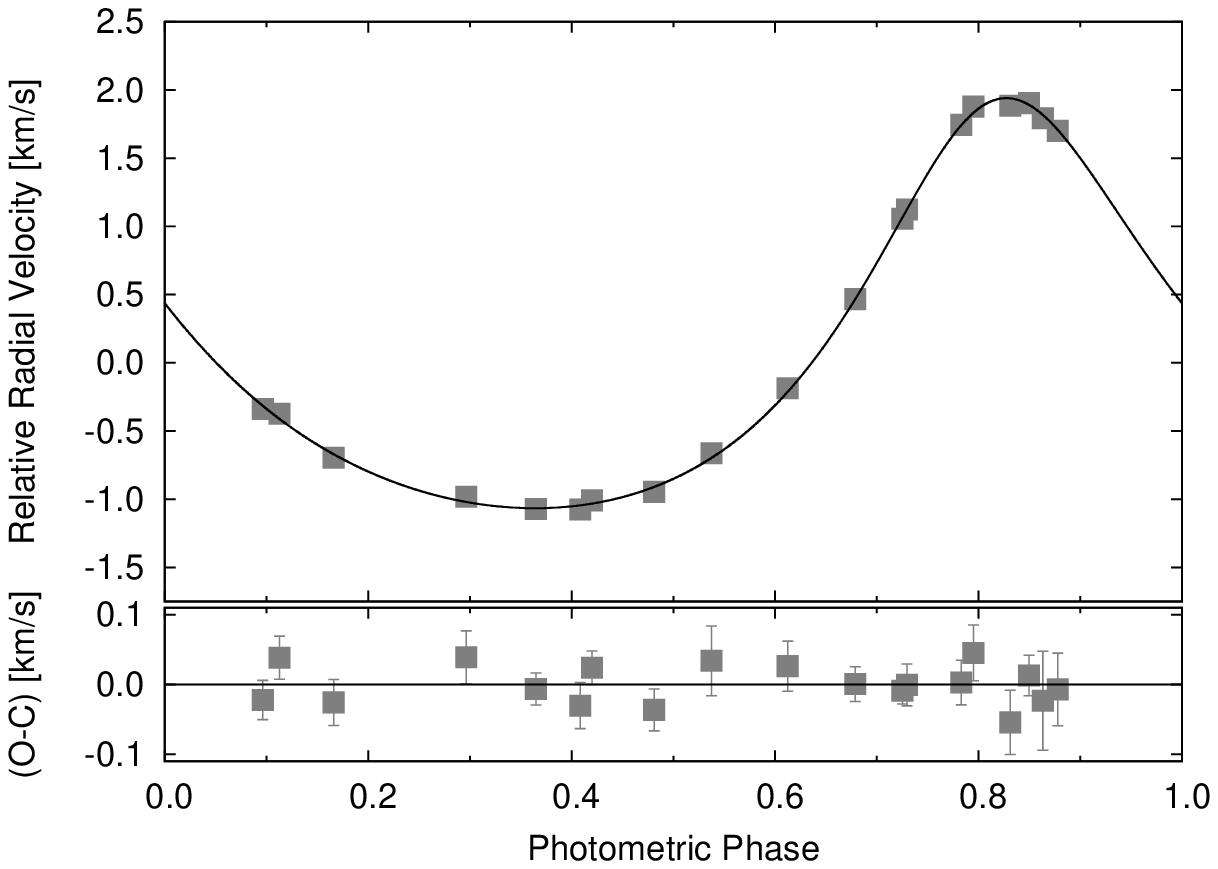}
\includegraphics[width=57mm,height=57mm]{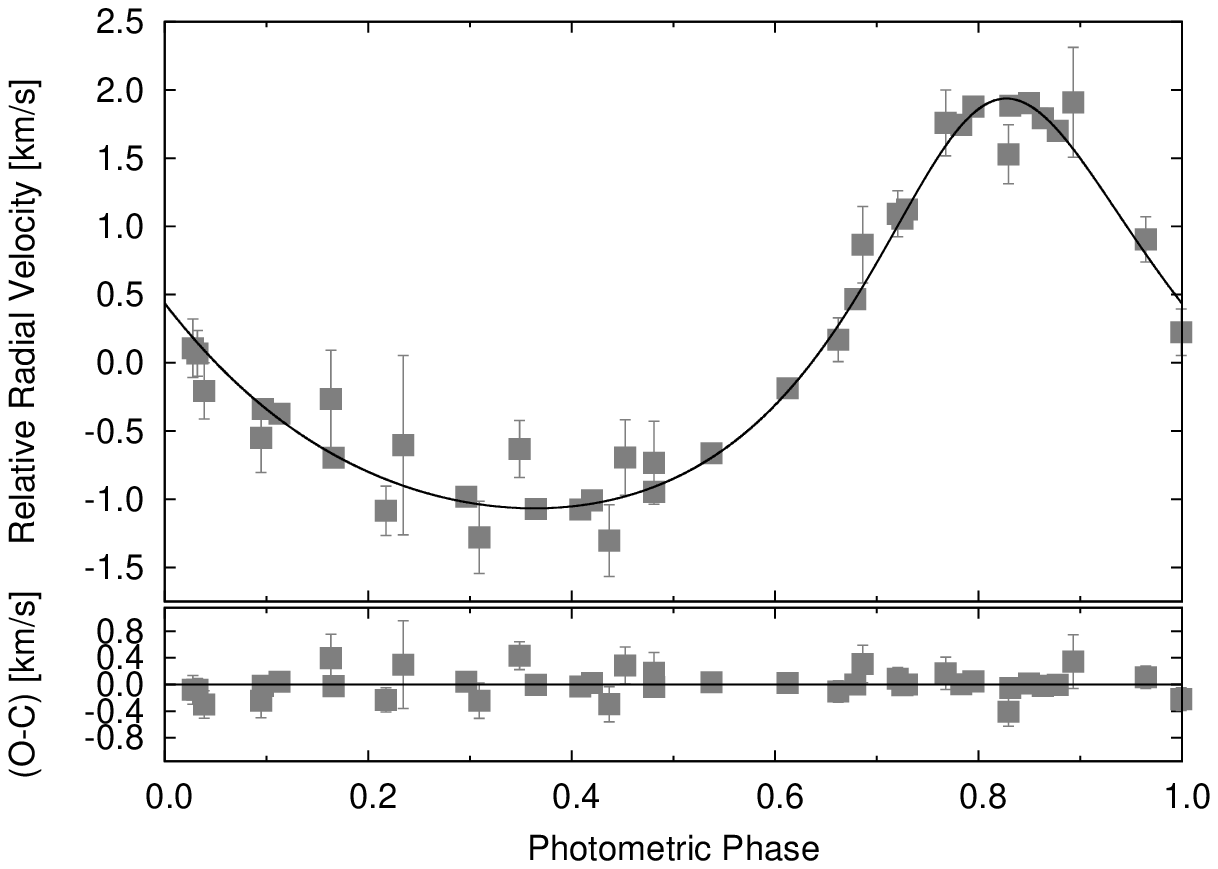}
}   
\caption{RV measurements of XO-3 phased with the orbital period of $P_\mathrm{orb}=3.191523$ days. The zero point in phase corresponds to the epoch of midtransit $T_0=2454449.8681$ HJD. The RV measurements are overplotted with our best JKTEBOP fit model. Corresponding residuals are also shown (bottom). The model was calculated based on simultaneous fit to the photometric data and RV data obtained at G1 (left panel), at OHP (middle panel) and at the all of here mentioned observatories (right panel). Note that RV measurements obtained at OHP have too small uncertainties to discern.}
\label{XO-RVFIT}
\end{figure*}

\begin{table}[b!]
\centering
\caption{Physical and orbital parameters of the XO-3 system summarized from papers Winn et al. 2008 [W8], Winn et al. (2009) [W9], H\'ebrard et al. 2008 [H8] and Wong et al. (2014) [W14].}
\label{XO-DATA}
\begin{tabular}{lll}
\hline
\hline
Parameter					& Value			& Ref      \\
\hline
\hline
Zero transit time $T_0$ [red. HJD]$^{a}$	& 54449.8681(2)		& [W8]     \\
Orbital period $P_\mathrm{orb}$ [d]		& 3.191523(6)		& [W8]     \\
Normalized semimajor axis $a/R_\mathrm{s}$	& 7.0(3)		& [W8]	   \\
Ratio of the radii $R_\mathrm{p}/R_\mathrm{s}$	& 0.0905(5)		& [W8]     \\
Inclination $i$ [deg] 	 			& 84.2(5)		& [W8]     \\
Eccentricity $e$				& 0.2769(17)		& [W14]    \\
Periastron longitude $\omega$ [deg]		& 346.3(1.3)		& [W9]     \\
Systemic velocity $\gamma$ [m.s$^{-1}$]		& -12045(8)		& [W9]     \\
RV semi-amplitude $K$ [m.s$^{-1}$]		& 1503(10)		& [H8]     \\   
Metallicity (Fe/H)				& -0.17(8)		& [W8]     \\
Effective temperature $T_\mathrm{eff}$ [K]	& 6429(100)		& [W8]     \\
Star surface gravity $\log g$			& 4.24(4)		& [W8]     \\ 
\hline          
\end{tabular}
\raggedright{\scriptsize{$^{a}$red. HJD = HJD $-$ 2~400~000}}
\end{table} 

In order to measure the orbital elements and the mass ratio of the system, Johns-Krull et al. (2008) as first obtained spectra of XO-3 at the 2.7m Harlan J. Smith Telescope and on the 11m Hobby-Eberly Telescope. The RV curve of XO-3 system by Johns-Krull et al. (2008) traces out an eccentric orbit with a RV semi-amplitude about 1471 m.s$^{-1}$. Since XO-3b is unusual in many respects (high mass, eccentric orbit and short orbital period), raising the possibility of the planet-planet scattering or Kozai oscillations, it was a subject of investigation concerning spin-orbit misalignment. H\'ebrard et al. (2008) observed the XO-3 system during the transit on night of January 28, 2008 with the SOPHIE instrument at the 1.93m telescope at OHP and reported the value of 70(15) [deg] as the sky-projected angle between the planetary orbital axis and the stellar rotation axis. Subsequently, Winn et al. (2009) reported the value of 37.3(3.7) [deg] based on observations with the Keck I 10m telescope. The origin of the $\sim 2\sigma$ difference is unclear.  

\begin{table*}[t!]
\centering
\caption{An overview of the best JKTEBOP fit parameters of the HAT-P-2, WASP-14 and XO-3 systems, resulting from the JKTEBOP fit to all the photometric data per object and RV data per object from selected observatories. Fixed parameters are listed without errors. Parameters from papers by Bakos et al. (2007) [B07], P\'al et al. (2010) [P10], Lewis et al. (2013) [L13], Joshi et al. (2009) [J09], Husnoo et al. (2011) [H11], Wong et al. (2015) [W15], Winn et al. (2008) [W08], Winn et al. (2009) [W09], H\'ebrard et al. (2008) [H08], Wong et al. (2014) [W14] and Claret \& Bloemen (2011) [C11] are summarized in 2nd column.}
\label{ALL-DATA}
\begin{tabular}{llllll}
\hline
\hline
Parameter						& Value			& Lit	& Value		& Value		& Value    	\\
							& from Lit		& Ref	& from G1	& from GAO+PO	& from G1+GAO+PO\\
\hline
\hline
\multicolumn{6}{c}{HAT-P-2b}\\
\hline
Zero transit time $T_0$ [HJD $-$ 2~400~000]		& 54387.4937(7)		& [P10] & 54387.4937	& 54387.4937    & 54387.4937    \\
Orbital period $P_\mathrm{orb}$ [d]			& 5.633472(6)		& [P10] & 5.633472	& 5.633472	& 5.633472      \\ 
Normalized semimajor axis $a/R_\mathrm{s}$		& 8.9(3)		& [P10]	& -- 		& --		& --	 	\\
Sum of fractional radii $(R_\mathrm{s}+R_\mathrm{p})/a$	& --			& --	& 0.143(11)	& 0.130(6)	& 0.119(5)	\\
Ratio of the radii $R_\mathrm{p}/R_\mathrm{s}$		& 0.0722(6)		& [P10] & 0.0677(2)	& 0.0678(2)	& 0.0678(2)     \\
Inclination $i$ [deg] 	 				& 86.7(8)		& [P10] & 86.4(4)	& 86.8(3)	& 87.2(3)	\\	  		
Eccentricity $e$					& 0.50910(48)		& [L13] & 0.58(4)	& 0.56(2)	& 0.510(16)	\\  		
Periastron longitude $\omega$ [deg]			& 188.09(39)		& [L13] & 165(6)	& 175(5)	& 181(5)	\\
Systemic velocity $\gamma$ [m.s$^{-1}$]			& $-$278(20)		& [B07] & $-$279(44)	& $-$197(31)	& $-$220(22) 	\\
RV semi-amplitude $K$ [m.s$^{-1}$]			& 983(17)		& [P10] & 1131(179)	& 904(52)	& 930(42)       \\
Linear LD coefficient ($z$-band)			& 0.186			& [C11]	& 0.186		& 0.186		& 0.186		\\
Non-linear LD coefficient ($z$-band)			& 0.302			& [C11]	& 0.302		& 0.302		& 0.302		\\
Light scale factor $L_\mathrm{sf}$ [mag]		& --			& --	& 7.87965(2)	& 7.87965(2)	& 7.87965(2)	\\
\hline
\hline          
Parameter						& Value			& Lit	& Value		& Value		& Value    	\\
							& from Lit		& Ref	& from G1	& from NOT+OHP	& from G1+NOT+OHP\\
\hline
\hline
\multicolumn{6}{c}{WASP-14b}\\
\hline
Zero transit time $T_0$ [HJD $-$ 2~400~000]		& 54463.5758(5)		& [J09]	& 54463.5758	& 54463.5758 	& 54463.5758	\\
Orbital period $P_\mathrm{orb}$ [d]			& 2.243752(10)		& [J09]	& 2.2437671(3)	& 2.2437669(2)	& 2.2437669(2)	\\ 
Orbital semimajor axis $a$ [AU]				& 0.0360(10)		& [J09]	& --		& --		& --		\\
Stellar radius $R_\mathrm{s}$ [R$_{\odot}$]		& 1.30(6)		& [J09] & --		& --		& --		\\
Transit depth $(R_\mathrm{p}/R_\mathrm{s})^2$ [mag]	& 0.0102(2)		& [J09]	& --		& --		& --		\\
Sum of fractional radii $(R_\mathrm{s}+R_\mathrm{p})/a$	& --			& --	& 0.180(12)	& 0.171(6)	& 0.171(6)	\\
Ratio of the radii $R_\mathrm{p}/R_\mathrm{s}$		& -- 			& -- 	& 0.0973(8)	& 0.0972(8)	& 0.0972(8)	\\
Inclination $i$ [deg] 	 				& 84.3(6)		& [J09]	& 85.1(8)	& 85.6(5)	& 85.6(5)	\\	  		
$k=e~\cos \omega$					& $-$0.02474(78)	& [W15]	& $-$0.02(3)	& $-$0.0254(9)	& $-$0.0253(9)	\\  		
$h=e~\sin \omega$					& $-$0.0792(31)		& [W15]	& $-$0.03(6)	& $-$0.085(2)	& $-$0.085(2)	\\
Systemic velocity $\gamma$ [m.s$^{-1}$]			& $-$4985(3)		& [H11]	& $-$4985(48)	& $-$4985.0(1.7)& $-$4985.0(1.7)\\
RV semi-amplitude $K$ [m.s$^{-1}$]			& 991(3)		& [H11]	& 1041(54)	& 993(3)	& 993(3)	\\
Linear LD coefficient ($R$-band)			& 0.267			& [C11]	& 0.267		& 0.267		& 0.267		\\
Non-linear LD coefficient ($R$-band)			& 0.324			& [C11]	& 0.324		& 0.324		& 0.324		\\
Light scale factor $L_\mathrm{sf}$ [mag]		& --			& --	& $-$0.00521(8)	& $-$0.00521(8)	& $-$0.00521(7)	\\
\hline
\hline
Parameter						& Value			& Lit	& Value		& Value		& Value    	\\
							& from Lit		& Ref	& from G1	& from OHP	& from G1+OHP	\\
\hline
\hline
\multicolumn{6}{c}{XO-3b}\\
\hline
Zero transit time $T_0$ [HJD $-$ 2~400~000]		& 54449.8681(2)		& [W08] & 54449.8681	& 54449.8681    & 54449.8681	\\
Orbital period $P_\mathrm{orb}$ [d]			& 3.191523(6)		& [W08] & 3.191523	& 3.191523	& 3.191523      \\ 
Normalized semimajor axis $a/R_\mathrm{s}$		& 7.0(3)		& [W08] & -- 		& --		& --	 	\\
Sum of fractional radii $(R_\mathrm{s}+R_\mathrm{p})/a$	& --			& --	& 0.159(13)	& 0.143(9)	& 0.144(8)	\\
Ratio of the radii $R_\mathrm{p}/R_\mathrm{s}$		& 0.0905(5)		& [W08] & 0.1012(19)	& 0.1005(19)	& 0.1005(19)	\\
Inclination $i$ [deg] 	 				& 84.2(5)		& [W08] & 84.0(9)	& 85.1(7)	& 85.1(7)	\\	  		
Eccentricity $e$					& 0.2769(17)		& [W14] & 0.255(19)	& 0.285(3) 	& 0.285(3)	\\  		
Periastron longitude $\omega$ [deg]			& 346.3(1.3)		& [W09] & 346(13)	& 348.8(1.5)	& 348.9(1.5)	\\
Systemic velocity $\gamma$ [m.s$^{-1}$]			& $-$12045(8)		& [W09] & $-$12045(51)	& $-$12030(7)	& $-$12030(7) 	\\
RV semi-amplitude $K$ [m.s$^{-1}$]			& 1503(10)		& [H08] & 1441(89)	& 1502(10)	& 1501(10)      \\
Linear LD coefficient ($V$-band)			& 0.351			& [C11]	& 0.351		& 0.351		& 0.351		\\
Non-linear LD coefficient ($V$-band)			& 0.311			& [C11]	& 0.311		& 0.311		& 0.311		\\
Light scale factor $L_\mathrm{sf}$ [mag]		& --			& -- 	& $-$0.0001(2)	& $-$0.0001(2)	& $-$0.0001(2)	\\
\hline
\hline          
\end{tabular}
\end{table*} 

The semi-amplitude of the RV is $K=1503(10)$ m.s$^{-1}$ (H\'ebrard et al. 2008), therefore it is the easiest detectable planet in our sample for our echelle spectrograph, mounted on the 0.6m telescope. On the other hand, XO-3b has a fainter parent star than HAT-P-2b or WASP-14b. Together we worked with 39 (20+19) RV measurements (see Table \ref{XO-RV}). The averaged scatter of G1 RVs is about 260 m.s$^{-1}$. First, we fitted only our G1 RV data, simultaneously with the photometric data (Figure \ref{XO-LC}). Initial parameters for the JKTEBOP fitting were taken from published papers by Winn et al. 2008, 2009; H\'ebrard et al. 2008 and Wong et al. (2014). These are summarized in Table \ref{XO-DATA}. The sum of fractional radii ($(R_\mathrm{s}+R_\mathrm{p})/a$) was calculated from the normalized semi-major axis ($a/R_\mathrm{s}$) and from the ratio of the radii ($R_\mathrm{p}/R_\mathrm{s}$). The interpolated quadratic LD coefficients for the $V$-band were 0.351 and 0.311. The fitted RV data are shown in Figure \ref{XO-RVFIT} -- left panel and the best fit parameters are summarized in Table \ref{ALL-DATA}. Subsequently, we fitted OHP RV data, simultaneously with the photometric data. The fitted RV data are shown in Figure \ref{XO-RVFIT} -- middle panel and the best fit parameters are summarized in Table \ref{ALL-DATA}. The averaged RV scatter of OHP data is about 35 m.s$^{-1}$. The accuracy is sometimes better than 30 m.s$^{-1}$. At G1 RVs in a few cases we achieved accuracy better than 200 m.s$^{-1}$. The best accuracy at OHP RVs is 19 m.s$^{-1}$, at G1 RVs it is 159 m.s$^{-1}$. As the final step, we simultaneously fitted photometric and all RV data. The fitted RV data are shown in Figure \ref{XO-RVFIT} -- right panel and the best fit parameters are summarized in Table \ref{ALL-DATA}. 
   
As we have seen previously, Figure \ref{XO-RVFIT} shows again that the JKTEBOP fit model derived from G1 RV values is similar to the JKTEBOP fit model derived from values adopted from H\'ebrard et al. (2008). We can compare e.g., the periastron longitude $\omega$ ($346(13)$ versus $348.8(1.5)$ ~[deg]), the systemic velocity $\gamma$ ($-12045(51)$ versus $-12030(7)$ ~[m.s$^{-1}$]), or the orbit eccentricity $e$ ($0.255(19)$ versus $0.285(3)$). The parameter values derived from our G1 RVs are, however, more uncertain. The best fit parameters resulting from G1 RVs are also, in general, consistent with literature values. It is seen e.g., in the case of the periastron longitude $\omega$ ($346.3(1.3)$ versus $346(13)$ ~[deg]), or in the case of the systemic velocity $\gamma$ ($-12045(8)$ versus $-12045(51)$ ~[m.s$^{-1}$]). We can see again that literature values are given with better accuracy. We did not confirm only the ratio of the radii $R_\mathrm{p}/R_\mathrm{s}$ ($0.0905(5)$ versus $0.1012(19)$; $\sim 5.6 \sigma$). The scatter of the RV data points from the fit are also very different: $\pm$500 m.s$^{-1}$ in the case of the G1 RV data and $\pm$50 m.s$^{-1}$ in the case of the OHP RV data.  

\section{Conclusions}
\label{cnc}

In this paper we described in details the RV analysis of the first three transiting planetary systems, HAT-P-2, WASP-14 and XO-3, observed during the years 2014 -- 2016 at the Star\'a Lesn\'a Observatory -- G1 with the off-shelf low-cost eShel spectrograph, made by the French company Shelyak, and mounted on a 0.6m telescope. The main scientific goal of the study was to compare the precision of our G1 RV measurements with precision of RV data achieved with echelle spectrographs of other sub-meter-, meter- and two-meter-class telescopes. We also investigated the applicability of our RV data for modeling orbital parameters.

The expected RV precision (neglecting instrumental effects such as temperature- and pressure-related deformations or slit effects) can be estimated using formula of Hatzes, Cochran \& Endl (2010):

\begin{equation}
\sigma_{\rm RV} = \frac{C1 ~ v \sin i}{SNR ~ \sqrt{R^3 B f}},
\label{hatzes}
\end{equation}

\noindent where SNR is the signal-to-noise ratio of the data, $R$ is the resolving power of the spectrograph, $B$ is the wavelength coverage in \AA, $v \sin i$ is the projected rotational velocity of the star in km.s$^{-1}$ and the dimensionless function $f = f$(sp. type) reflects the line density. The proportionality constant was estimated by Hatzes, Cochran \& Endl (2010) as $C1 = 2.4 \times 10^{11}$ \AA$^{1/2}$. For stars with spectral type F the authors give $f \sim 1$. If we compare expected precision for the same star obtained with different spectrographs (resolution $R$) and telescopes (diameter $D$), assuming the same exposure time, instrument throughtput and atmospheric extinction we have:

\begin{equation}
\sigma_{\rm RV} = \frac{C2}{D \sqrt{R^3}}.
\label{}
\end{equation}
   
\begin{figure}[t!]
\centering
\includegraphics[width=80mm]{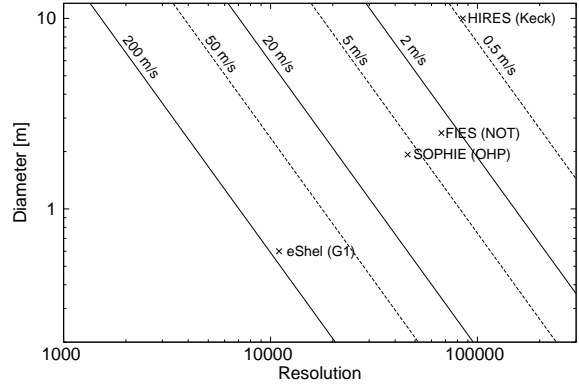}   
\caption{Expected RV precision as a function of the telescope diameter and the spectrograph resolution. The precision is scaled to match the precision achieved for HAT-P-2 with eShel on the 0.6m telescope at Star\'a Lesn\'a. The precision curves assume the same exposure time, telescope-spectrograph throughput and the same object.}
\label{resolve}
\end{figure}

\noindent The constant $C2 \sim 1.2 \times 10^8$ m$^2$.s$^{-1}$ was found to match the precision for HAT-P-2 RV data with eShel on our 0.6m telescope. The resulting precision dependence is shown for several spectrographs used to observe HAT-P-2, WASP-14 and XO-3 in Figure \ref{resolve}. The expected precision is reasonably matching observed uncertainty of other spectrographs. Somewhat higher precision of other spectrographs compared to eShel is caused by much longer exposures (typically three 900-sec exposures were combined) of our data used to determine $C2$.

Based on our results, we can conclude that the spectrograph is a useful instrument for study of objects with relatively small RV amplitude. We achieved an averaged precision of about 170 m.s$^{-1}$ in the case of HAT-P-2, 220 m.s$^{-1}$ in the case of WASP-14 and 260 m.s$^{-1}$ in the case of XO-3 system. These values are sufficient for exoplanet RV detections and spectroscopic follow-up measurements of massive exoplanets on close-in orbits. The accuracy is well comparable with the averaged RV scatter achieved with other sub-meter- and meter-class telescopes, e.g., 170 versus 177 m.s$^{-1}$ in the case of HAT-P-2 RV data, obtained at G1, GAO and PO with a 0.6m, 0.5m and 1m telescope, respectively. In comparison with two-meter-class telescopes, our instrumentation gives a RV scatter about one order greater, e.g., 260 versus 35 m.s$^{-1}$ in the case of XO-3 data, obtained at G1 and OHP with a 0.6m and 1.93m telescope, respectively. This difference is primarily caused by the telescope diameter size, but it also depends on many properties of instrumentation which determine the RV stability (see Pribulla et al. 2015). From this aspect, the spectrograph thermal stabilization is a key point for the RV accuracy. SOPHIE at OHP is installed in a thermally-controlled and isolated box with a daily thermal stability better than 0.01 $^{\circ}$C inside a two-stage thermalised room (Perruchot et al. 2008). FIES at NOT is also located in a protected building at a stable temperature within 0.02 $^{\circ}$C (Telting et al. 2014). Our spectrograph itself at G1 is located in the cellar below the dome, where is a quasi stable temperature, however, the room is without thermal-control. The RV precision is also limited by the method of spectral calibration. To obtain the best calibration properties we can acquire ThAr spectra just before and after the object spectra recording and we have to use the same current of the hollow-cathode lamp. FIES and SOPHIE, however, offer the option of simultaneous observations of object and reference ThAr spectra. In addition they allow to analyze the sky background or the moonlight contamination simultaneously as well. Moreover, RV accuracy of our data is limited by precision of the dispersion solution, which is 30-50 m.s$^{-1}$ for our instrumentation and which mainly depends on the spectral resolution (Pribulla et al. 2015). This is why the further improvement of the RV system of the eShel spectrograph at G1 cannot be achieved by providing more stable environment nor by using a larger telescope.
    
On the other hand, our best JKTEBOP fit results show that RV data, obtained with our instrumentation, can be used to determine orbital parameters of massive close-in exoplanets. The scatter of the RV data points from the model were in all cases up to $\pm$500 m.s$^{-1}$. In general, we can conclude that best fit parameters, resulting from the G1 RV data are in good agreement with published parameters. Literature parameter values are, however, given with better accuracy. Furthermore, in comparison with NOT/OHP RV data, due to the relatively lower RV accuracy of G1 RV data, we can determine system parameters with bigger error interval only. This is also the reason, why parameters derived from NOT/OHP RV data and from combined G1 and NOT/OHP observations are very similar (see Table \ref{ALL-DATA}). Since data obtained at G1 have lower accuracy, these are low-weighted during the model fitting procedure when we combined G1 and NOT/OHP observations and have minimal influence on parameter determination.       

In addition, there is an educational point of view of our echelle spectrograph for students on different level of study. The instrument can be used as a didactic tool in teaching of spectroscopy.  

\vspace{0.5cm}
\acknowledgements
The authors thank V. Koll\'{a}r for his technical assistance, B. Cs\'ak for the RV data and J. Budaj for his comments and discussion. This work has been supported by the VEGA grant of the Slovak Academy of Sciences No. 2/0143/14, by the Slovak Research and Development Agency under the contract No. APVV-0158-11 and by the realization of the Project ITMS No. 26220120009, based on the Supporting Operational Research and Development Program financed from the European Regional Development Fund. MS thanks Piezosystem Jena for financial support.

\end{document}